\documentclass[prd]{revtex4}
\usepackage{graphicx}

\newcommand{\tr}{{\rm tr}}

\newcommand{\dslash}{{\partial \hskip -6pt /}}

\begin{document}

\title{Thermal Fluctuations of Induced Fermion Number}

\author{Gerald V. Dunne and Kumar Rao}

\affiliation{Department of Physics, University of Connecticut, Storrs CT
06269-3046, USA\\}


\begin{abstract}
We analyze the phemomenon of induced fermion number at finite temperature. At 
finite temperature, the induced fermion number $\langle N\rangle$ is a thermal
expectation value, and we compute the finite temperature fluctuations, 
$(\Delta N)^2=\langle N^2\rangle-\langle N\rangle^2$.
While the zero temperature induced fermion number is topological and is a
sharp observable, the finite temperature induced fermion
number is generically nontopological, and is not a sharp observable.  The
fluctuations are due to the mixing of states inherent in any finite
temperature expectation value. We analyze in detail two different cases in
$1+1$ dimensional field theory: fermions in a kink background, and fermions
in a chiral sigma model background. At zero temperature the induced fermion
numbers for these two cases are very similar, but at finite temperature they
are very different. The sigma model case is generic and the induced fermion
number is nontopological, but the kink case is special and the fermion number
is topological, even at finite temperature. There is a simple physical
interpretation of all these results in terms of the spectrum of the fermions
in the relevant background, and many of the results generalize to higher
dimensional models.

\end{abstract}
\maketitle

\section{Introduction}

The phenomenon of induced fermion number arises due to the interaction of
fermions with nontrivial topological backgrounds (e.g., solitons, vortices,
monopoles, skyrmions), and has many applications ranging from polymer physics
to particle physics \cite{jr,gw,jackiw2,niemi,polymer,roman}. The original
fractional fermion number result of Jackiw and Rebbi \cite{jr} provides
the physical explanation of the existence of spinless charged excitations
in polymers \cite{polymer,roman,ssh,rice,js}. The adiabatic analysis of
Goldstone and Wilczek \cite{gw} in systems without conjugation symmetry
has important implications for model field theories in particle physics,
such as bag models \cite{bag}, electroweak theories \cite{eric}, and
chiral sigma models \cite{diakonov}. At zero temperature, the induced
fermion  number is a topological quantity, and is related to the spectral
asymmetry of the relevant Dirac operator, which counts the difference
between the number of positive and negative energy states in the fermion
spectrum \cite{jackiw2,niemi,roman,rich,wilczek,yamagishi,boy,mike,manu,jaffe2}.
Mathematical results, such as index theorems \cite{manu,niemi} and
Levinson's theorem
\cite{yamagishi,boy,mike,jaffe2}, imply that the zero temperature
induced fermion number is determined by the asymptotic topological properties
of the background fields. This topological character of the induced fermion
number is a key feature of its application in certain model field theories
\cite{bag,eric,diakonov}.

At finite  temperature, the situation is less clear-cut. In
several examples, namely the $1+1$ dimensional chiral kink
background \cite{ns,soni,keil} and the $3+1$ dimensional Dirac
\cite{cp,monopole} and 't Hooft-Polyakov monopole \cite{goldhaber}
backgrounds, the finite temperature induced fermion number has been shown to
be temperature dependent, but still topological in the sense that the only
dependence on the background field is through its asymptotic properties. 
However, it has recently been demonstrated \cite{ad} that the finite temperature
induced fermion number need not be topological. Explicitly, in a
$1+1$ dimensional chiral sigma model, the finite temperature induced fermion
number depends on the detailed structure of the background, not just its
asymptotic behavior. This result corrects several previous analyses
\cite{midorikawa,nnp} that had claimed that the finite temperature fermion
number was in general a topological quantity (at zero chemical potential). In
\cite{ad}, an explicit calculation showed that the finite temperature fermion
number can be nontopological, and a simple physical explanation was given for
the origin of the nontopological fermion number as the plasma response of the
fermions to the inhomogeneous background. A possible source of confusion here
is that while the kink and sigma model cases are very similar at zero
temperature, it is not widely appreciated that at finite temperature they are
very different. In this paper we present more details of this analysis, and
we present a general argument that the finite temperature fermion number
naturally separates into a temperature-independent topological piece that
corresponds to vacuum polarization effects, and a temperature-dependent piece
that is {\it generically} nontopological, and which corresponds to the
thermal occupation of excited states of the fermion Fock space. While this
temperature-dependent piece is generically nontopological, for certain
particular backgrounds (such as, for example, the kink background) the fermion
spectrum has a special symmetry, which has the consequence that the
temperature-dependent corrections are in fact themselves topological.

Another motivation for our work concerns the nature of the induced fermion
number as a sharp quantum observable. At zero temperature it has been shown
that the fractional induced fermion number is indeed a sharp observable
\cite{kivelson,bell,jackiw,frishman,charge}. In this paper we address this
question at finite temperature by computing the rms fluctuations, $(\Delta
N)^2=\langle N^2\rangle-\langle N\rangle^2$, in the induced fermion number
$\langle N\rangle$. We compute the finite temperature fluctuation in two
different cases, and show that $\Delta N$ is generally nonzero (and
nontopological) at finite temperature, but that $\Delta N$ vanishes at zero
temperature. This indicates that
$N$ is {\it not} a sharp quantum observable at nonzero
temperature. The nonvanishing fluctuations are due to the mixing of Fock
states inherent in the thermal expectation value. It is interesting to note
that another example of non-sharp fractional fermion number (but not in the
context of temperature dependence) has been discussed recently in the context
of liquid helium bubbles \cite{helium}.

Our analysis of finite temperature fermion number has also been motivated
by the realization over recent years that certain types of anomalies in
zero temperature field theory become much more subtle at finite temperature.
For example, Pisarski et al have explained \cite{pisarski} why finite
temperature anomalous $\pi^0$ decay amplitudes are temperature dependent even
though the chiral anomaly (a topological object that is known to be related
to the anomalous $\pi^0$ decay amplitude at $T=0$) is known to be temperature
independent. These and related issues have also been explored in $1+1$
dimensions \cite{gelis}. And in odd spacetime dimensions, it has recently
been realized \cite{dll,deser,fosco,lh} that while the topological
Chern-Simons term is the only parity-violating term that can be induced in
the $T=0$ effective action, at finite temperature there are infinitely many
other parity-violating terms, all of which vanish identically at zero
temperature, but all of which are crucial, for example, for understanding how
it is possible to maintain large gauge invariance at finite temperature.
These results have led us to reconsider the related general question of
induced fermion number at finite temperature.

In Section II we define what is meant by finite temperature induced fermion
number, and indicate how it can be computed. In Section III we give more
details of the derivations and results of \cite{ad} for the finite temperature
induced fermion number $\langle N\rangle$ for a kink background and for a
sigma model background in $1+1$ dimensional field theory. Section IV presents
the computation of the finite temperature fluctuations, 
$( \Delta N)^2=\langle N^2\rangle-\langle N\rangle^2$, in the induced fermion
number for the kink case and for the sigma model case. In Section V we give a
simple physical interpretation of all these results in terms of the thermal
occupation of the available levels in the Dirac spectrum, according to
Fermi-Dirac statistics. We conclude in Section VI and give some general
comments regarding the extension of our results to higher dimensional
theories.

\section{Finite Temperature Induced Fermion Number}

Consider an abelian model in $1+1$ dimensions with fermions interacting via
scalar and pseudoscalar couplings to two bosonic fields $\phi_1$ and $\phi_2$.
For the purposes of this paper, $\phi_1$ and $\phi_2$ will be considered as
static classical background fields. The Lagrangian is
\begin{equation}
{\cal L}=i\,\bar{\psi}\dslash\,\psi -\bar{\psi}\left(\phi_1+i\,
 \gamma_5\,\phi_2\right)\psi
\label{lag}
\end{equation}
We will concentrate on two important physical cases:

\begin{figure}
\includegraphics*{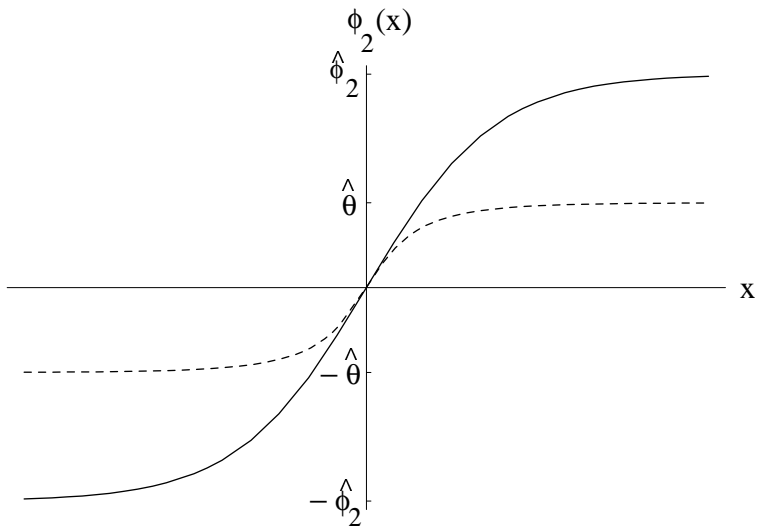}
\caption{In the kink case (\protect{\ref{kink}}), the kink function
$\phi_2(x)$ has the form shown in the solid line, with asymptotic
values $\pm\hat{\phi_2}$. The corresponding angular field defined in
(\protect{\ref{theta}}), 
$\theta(x)={\rm arctan}(\frac{\phi_2(x)}{m})$, also has a kink-like
shape, as shown in the dashed line, with asymptotic values 
$\-pm\hat{\theta}$, where $\hat{\theta}= {\rm arctan}(\frac{\hat{\phi_2}}{m})$.
Note that in the kink case, the allowed values of $\hat{\theta}$ lie in 
the first branch $[-\frac{\pi}{2},\frac{\pi}{2}]$ of the arctan function.}
\label{f1}
\end{figure}

\begin{figure}[b]
\includegraphics*{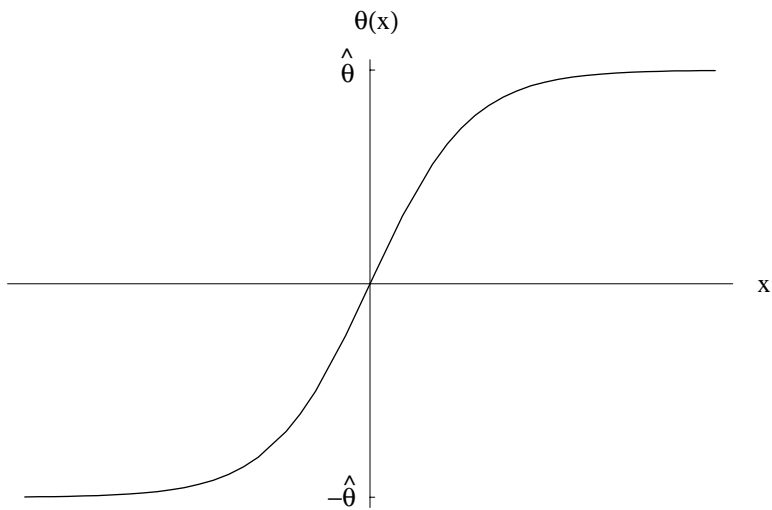}
\caption{In the sigma model case (\protect{\ref{sigma}}), it is the
angular field $\theta(x)$ defined in (\protect{\ref{theta}}) that has a
kink-like shape, with asymptotic
values $\pm\hat{\theta}$.}
\label{f2}
\end{figure}

(i) kink case \cite{jr}: $\phi_1$ is constant, but
$\phi_2=\phi_2(x)$ has a kink-like shape, as shown in Fig. 1:
\begin{equation}
\phi_1=m,\qquad {\rm and}  \quad
\phi_2={\rm kink},\,\, {\rm with}\quad\phi_2(\pm\infty)=\pm\hat{\phi_2}
\label{kink}
\end{equation}

(ii) sigma model case \cite{gw}: $\phi_1$ and $\phi_2$ are both x-dependent,
but are constrained to the ``chiral circle'':
\begin{equation}
\phi_1^2+\phi_2^2=m^2
\label{sigma}
\end{equation}

In each case, we define a corresponding angular field
\begin{eqnarray}
\theta(x)\equiv {\rm arctan}\,\left(\frac{\phi_2(x)}{\phi_1(x)}\right)
\label{theta}
\end{eqnarray}
In the kink case (\ref{kink}), the angular field $\theta(x)={\rm
arctan}\,(\frac{\phi_2(x)}{m})$ also has a kink-like shape,
as shown in Fig. 1. In the sigma model case (\ref{sigma}), the angular field
$\theta(x)$ has the interpretation of a local chiral angle, since the
chiral constraint (\ref{sigma}) allows us to write $\phi_1=m\cos\theta$ and
$\phi_2=m\sin\theta$, so that the interaction term in the
Lagrangian (\ref{lag}) is
\begin{equation}
m \bar{\psi}\left(
\cos\theta+i\gamma_5 \sin\theta\right)\psi=m \bar{\psi}\, e^{i\gamma_5
\theta}\, \psi
\label{exp}
\end{equation}
In this sigma model case, it is the angular field $\theta(x)$ that has a
kink-like shape, as shown in Fig. 2.

The second-quantized fermion number operator is defined \cite{bjorken} as 
$N=\frac{1}{2}[\Psi^\dagger, \Psi]$. The fermion field operator $\Psi$ may
be expanded in a complete set of eigenstates of the Dirac Hamiltonian $H$
for fermions in the presence of the background fields $\phi_1$ and $\phi_2$.
The Dirac  Hamiltonian corresponding to the Lagrangian (\ref{lag}) is
\begin{eqnarray}
H&=&-i\gamma^0\gamma^1\nabla +\gamma^0 \phi_1(x)+i\gamma^0\gamma_5
\phi_2(x)\nonumber\\
&=&\left(\matrix{\phi_1 & -i\nabla-i\phi_2\cr -i\nabla+\phi_2
&-\phi_1}\right)
\label{ham}
\end{eqnarray}
where $\nabla\equiv\frac{d}{dx}$, and we have chosen to work with the
Dirac matrices: $\gamma^0=\sigma_3$, $\gamma^1=i\sigma_2$, and
$\gamma^5=-\sigma_1$. The presence of the background fields modifies the
fermion spectrum from the free case. All necessary information 
about the fermion spectrum can be encoded in terms of the spectral function
$\sigma(E)$ of the Dirac Hamiltonian in (\ref{ham}):
\begin{eqnarray}
\sigma(E)=\frac{1}{\pi}\, {\cal I}m\, {\rm
Tr}\left(\frac{1}{H-E-i\epsilon}\right)
\label{spectral}
\end{eqnarray}
The zero temperature vacuum expectation value, 
$\langle 0|N|0\rangle\equiv\langle N\rangle_0$, of the number operator can
be expressed as \cite{jackiw2,niemi}
\begin{eqnarray}
\langle N\rangle_0=-\frac{1}{2}\int_{-\infty}^\infty dE\, \sigma(E)\,
{\rm sign}(E)
\label{asymmetry}
\end{eqnarray}
where the subscript $0$ refers to zero temperature. Thus,
$\langle N\rangle_0$ is expressed in terms of the {\it spectral asymmetry} of
the Dirac operator, which counts the number of positive energy states minus
the number of negative energy states. 

At zero temperature, in both the kink case (\ref{kink}) and the sigma model
case (\ref{sigma}), the induced fermion number (\ref{asymmetry}) is given by
the simple expression \cite{gw}:
\begin{equation}
\langle N\rangle_0=\frac{1}{2\pi}\int_{-\infty}^\infty dx\, \theta^\prime
=\frac{1}{\pi}\,\hat{\theta}
\label{nzero}
\end{equation}
where $\theta(x)$ is the angular field defined in (\ref{theta}), and 
$\hat{\theta}$ is the asymptotic value of $\theta(x)$ at $x=+\infty$.
The zero temperature fermion number $\langle N\rangle_0$ is {\it topological}
in the sense that it depends only on the asymptotic value
$\hat{\theta}$, and not on the detailed shape of $\theta(x)$. The original
conjugation symmetric case of Jackiw and Rebbi
\cite{jr} is obtained by taking $m\to 0$ in the kink case (\ref{kink}),
in which case $\langle N\rangle_0\to \pm\frac{1}{2}$. In Section IV we will
show that the zero temperature expectation value 
(\ref{asymmetry}) is in fact also a sharp eigenvalue.

At nonzero temperature, the induced fermion number is a thermal
expectation value 
\begin{eqnarray}
\langle N\rangle_{\rm T}&=& \frac{{\rm Tr}\,(N\, e^{-\beta H})}{{\rm Tr}\, 
(e^{-\beta H})}\nonumber\\
&=&-\frac{1}{2}\int_{-\infty}^\infty dE\, \sigma(E)\,\tanh\left(\frac{\beta
E}{2}\right)
\label{thermalexp}
\end{eqnarray}
where $\beta=\frac{1}{T}$ is the inverse temperature. Note that as $T\to
0$, which means $\beta\to\infty$, only the vacuum state survives in the
trace, and $\langle N\rangle_{\rm T}$ reduces to the vacuum expectation value
$\langle N\rangle_0$. Correspondingly,  $\tanh(\frac{\beta E}{2})\to {\rm
sign}(E)$, so that the integral expression in
(\ref{thermalexp}) reduces smoothly to the spectral asymmetry expression in
(\ref{asymmetry}). In fact, the finite temperature expression
(\ref{thermalexp}) provides a physically natural, and computationally
simple, regularization of the spectral asymmetry (\ref{asymmetry}). 
In Section IV we will show that at nonzero temperature the fermion number 
expectation value (\ref{thermalexp}) is not a sharp eigenvalue.

To compute $\langle N\rangle_0$ or $\langle N\rangle_{\rm T}$, one needs
information about the spectral function $\sigma(E)$. We stress, of course,
that $\sigma(E)$ has nothing to do with the temperature; it simply describes
the spectrum of the fermions in the presence of the static background fields
$\phi_1$ and $\phi_2$. One convenient way to
proceed is to use the expression (\ref{spectral}) for the spectral function
to write (\ref{thermalexp}) as a contour integral in the complex energy
plane:
\begin{equation}
\langle N\rangle_{\rm T}=-\frac{1}{2}\int_{\cal C}\,\frac{dz}{2\pi i}\,
\tr\left(\frac{1}{H-z}\right) {\rm tanh}\left(\frac{\beta z}{2}\right)
\label{contour}
\end{equation}
Here $\tr (\frac{1}{H-z})$ is 
the {\it resolvent} of the Dirac Hamiltonian $H$, and the contour 
${\cal C}$ is $(-\infty+i\epsilon,+\infty+i\epsilon)$ and 
$(+\infty-i\epsilon,-\infty-i\epsilon)$, as shown in Fig. 3. Since $H$
has a real spectrum, we can evaluate the contour integral (\ref{contour})
in two alternative ways. First, we can deform the contour around the
simple poles of the $\tanh(\frac{\beta z}{2})$ function, which occur
along the imaginary axis at the Matsubara modes:
$z_n=(2n+1)i\pi T$, for $n\in {\bf Z}$. This leads to an 
expression for $\langle N\rangle_{\rm T}$ as an infinite sum. Alternatively,
we can deform the $z$ contour around the poles and cuts of the spectrum of
$H$, which lie on the real axis. This leads to an integral representation for
$\langle N\rangle_{\rm T}$, which is just the Sommerfeld-Watson transform of
the infinite sum expression. We will see examples of these equivalent forms
below.

\begin{figure}
\includegraphics*{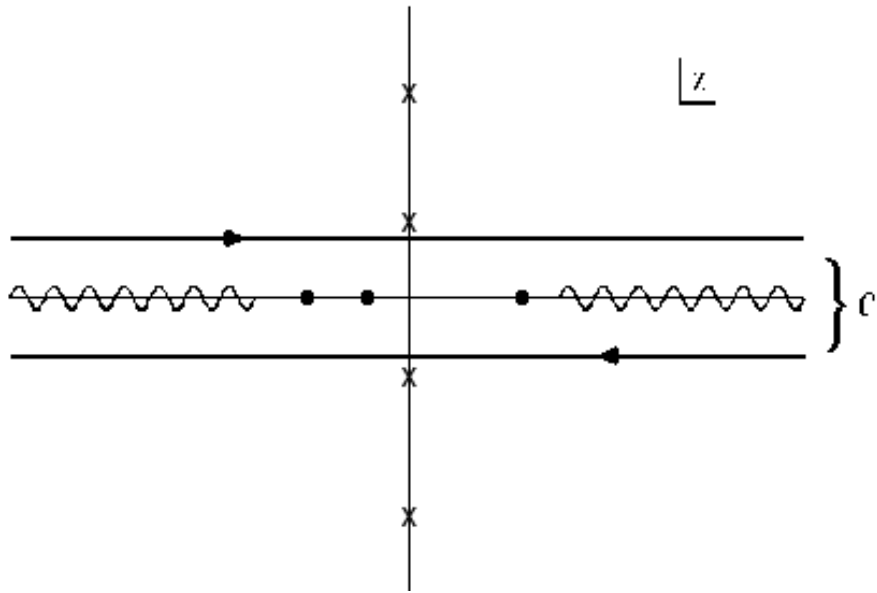}
\caption{The contour ${\cal C}$ in the complex energy plane used in the 
evaluation of the contour integral in (\protect{\ref{contour}}). The
spectrum of the Dirac Hamiltonian lies on the real axis and is indicated
schematically by the continuum cuts and some possible bound states in the
gap. The crosses on the imaginary axis depict the simple poles, $z_n=(2n+1)\pi i T$,
of $\tanh(\frac{\beta z}{2})$.}
\label{f3}
\end{figure}

\section{Results for Finite Temperature Induced Fermion Number}

Since $\tanh(\frac{\beta E}{2})$ is an odd function of $E$, it follows from
(\ref{thermalexp}) that to compute $\langle N\rangle_{\rm T}$ we only need the
{\it odd} part, $\sigma_{\rm odd}(E)=\frac{1}{2}[\sigma(E)-\sigma(-E)]$,
of the spectral function:
\begin{eqnarray}
\langle N\rangle_{\rm T}=-\int_0^\infty dE\,
\sigma_{\rm odd}(E)\,\tanh\left(\frac{\beta E}{2}\right)
\label{odd}
\end{eqnarray}
Correspondingly, we only need to know the {\it even} part of the
resolvent:
\begin{equation}
\langle N\rangle_{\rm T}=-\frac{1}{2}\int_{\cal C}\,\frac{dz}{2\pi i}\,
\left[\tr\left(\frac{1}{H-z}\right)\right]_{\rm even} {\rm
tanh}\left(\frac{\beta z}{2}\right)
\label{even}
\end{equation}
This fact simplifies the calculations considerably, and has important
physical consequences, as will be discussed in Section V.

\subsection{Finite T Fermion Number in the Kink Case}

\begin{figure}
\includegraphics*{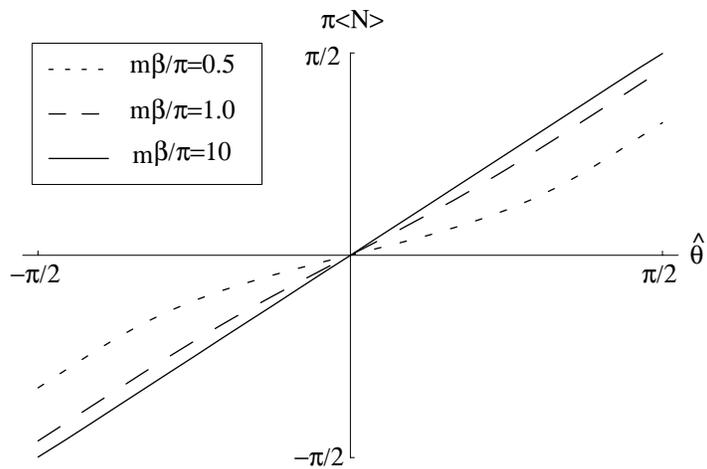}
\caption{Plots of ($\pi$ times) the finite temperature induced fermion number
$\langle N\rangle$, given in (\protect{\ref{kinkq}}) and
(\protect{\ref{kinkint}}), for the kink case. The plots show the dependence
of $\pi \langle N\rangle$ on the asymptotic value
$\hat{\theta}={\rm arctan}(\frac{\hat{\phi_2}}{m})$, for various values of
the temperature, as indicated. Note that for low temperature (large
$\beta$), $\langle N\rangle$ smoothly approaches the zero temperature result
$\langle N\rangle =\hat{\theta}/\pi$ in (\protect{\ref{nzero}}).}
\label{f4}
\end{figure}

\begin{figure}
\includegraphics*{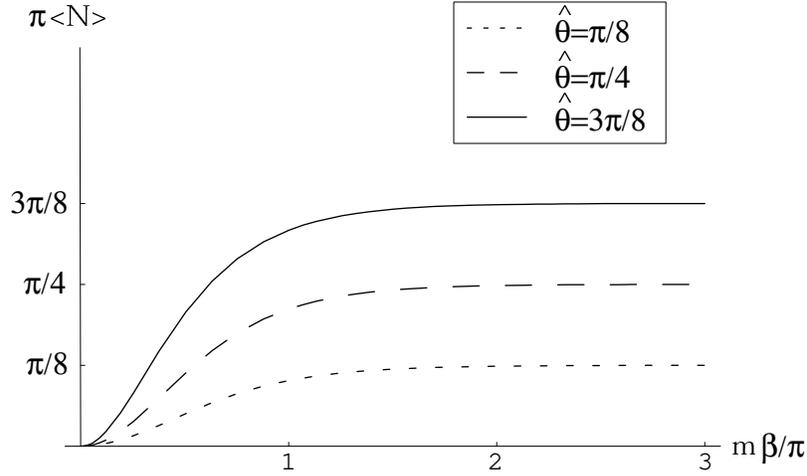}
\caption{Plots of ($\pi$ times) the finite temperature induced fermion number
$\langle N\rangle$, given in (\protect{\ref{kinkq}}) and
(\protect{\ref{kinkint}}), for the kink case. The plots
show the dependence of $\pi \langle N\rangle$ on the inverse temperature
$\frac{m\beta}{\pi}$, for various values of the asymptotic value
$\hat{\theta}$, as indicated. Note that for low temperature (large
$\beta$), $\langle N\rangle$ quickly saturates to the zero temperature value
$\pi\langle N\rangle =\hat{\theta}$. Also note that $\langle N\rangle$
vanishes as $T\to\infty$.}
\label{f5}
\end{figure}

For the kink case (\ref{kink}) we can use the remarkable result (a
special case of the Callias index theorem \cite{callias}) that
the even part of the resolvent (and hence the odd part of the spectral
function) can be computed exactly for {\it any} kink
$\phi_2(x)$:
\begin{eqnarray}
\left[\tr\left(\frac{1}{H-z}\right)\right]_{\rm even} 
={- m \hat{\phi}_2 \over (m^2-z^2)\,\sqrt{m^2+\hat{\phi}_2^2-z^2}}
\label{index}
\end{eqnarray}
Notice that this only depends on the kink $\phi_2(x)$ through its
asymptotic value $\hat{\phi}_2$; it is otherwise independent of the
details of the shape of the kink.

To motivate this fact,  we can express
the Dirac resolvent as $\tr\left(\frac{1}{H-z}\right)=
\tr\left((H+z)\frac{1}{H^2-z^2}\right)$, and note that, for
the kink case, $H^2$ takes the simple diagonal form 
\begin{equation}
H^2=\left(\matrix{-\nabla^2+m^2+\phi_2^2+\phi_2^\prime &0\cr 0&  
-\nabla^2+m^2+\phi_2^2-\phi_2^\prime}\right)
\label{hskink}
\end{equation}
Thus, 
\begin{eqnarray}
\left[\tr\left(\frac{1}{H-z}\right)\right]_{\rm even} 
&=& \tr\left({m\over -\nabla^2+\phi_2^2+\phi_2^\prime+m^2-z^2}\right)- 
\tr\left({m\over  -\nabla^2+\phi_2^2-\phi_2^\prime+m^2-z^2}\right)
\label{susy}
\end{eqnarray}
That is, we can express the {\it even} part of the Dirac resolvent as the
difference of two Schr\"odinger resolvents, with the corresponding
Schr\"odinger potentials, $V_\pm=\phi_2^2\pm\phi_2^\prime$, being {\it
isospectral}. This isospectrality property is the key ingredient for
proving (\ref{index}).

Given the result (\ref{index}), the induced fermion number (\ref{even})
for the kink case (\ref{kink}) is \cite{ad}
\begin{equation}
\langle N\rangle_{\rm T}= \frac{2}{\pi}\, \left(\frac{m\beta}{\pi}\right)^2
{\rm sin}\hat{\theta}\,
\sum_{n=0}^\infty {1\over \left((2n+1)^2 +(\frac{m\beta}{\pi})^2\right)
\sqrt{(2n+1)^2
\cos^2\hat{\theta}+(\frac{m\beta}{\pi})^2}}
\label{kinkq}
\end{equation}
where we recall that $\hat{\theta}\equiv{\rm
arctan}(\frac{\hat{\phi_2}}{m})$, and we can restrict our attention to
$-\frac{\pi}{2}\leq \hat{\theta}\leq \frac{\pi}{2}$. As an alternative to the
summation expression in (\ref{kinkq}), we can write $\langle N\rangle_{\rm
T}$ as an integral representation:
\begin{eqnarray}
\langle N\rangle_{\rm T}=\frac{1}{2}\, {\rm sign}(\hat{\theta})\,
\tanh(\frac{m\beta}{2})-\frac{\sin\hat{\theta}\,
\cos\hat{\theta}}{\pi}\int_1^\infty du\,
\frac{\tanh(\frac{m\beta}{2}\sec\hat{\theta}\, u)}{\sqrt{u^2-1}\,
(u^2-\cos^2\hat{\theta})}
\label{kinkint}
\end{eqnarray}
The kink case induced fermion number $\langle N\rangle_{\rm T}$ is plotted in
Fig. 4 as a function of $\hat{\theta}$ for various values of the temperature,
and in Fig. 5 as a function of $\frac{m\beta}{\pi}$, for various values of
$\hat{\theta}$. The integral form (\ref{kinkint}) makes the zero temperature
limit clear : using the integral 
\begin{eqnarray}
\int_1^\infty du\,
\frac{1}{\sqrt{u^2-1}\,
(u^2-\cos^2\hat{\theta})}=\frac{\frac{\pi}{2}-|\hat{\theta}|}{|\sin\hat{\theta}|
\cos\hat{\theta}}
\label{int}
\end{eqnarray}
it follows that the leading correction to $\langle N\rangle_0$ at low
temperature is exponentially small :
\begin{eqnarray}
\langle N\rangle_{\rm T}\,\sim\,\frac{\hat{\theta}}{\pi}- {\rm
sign}(\hat{\theta})\, e^{-m\beta}+\dots \quad ,
\quad \beta\to\infty
\label{leadingkink}
\end{eqnarray}
Thus $\langle N\rangle_{\rm T}$ reduces smoothly to the zero temperature
expression (\ref{nzero}). This can be seen clearly in Figs. 4 and
5. But at finite temperature, the expressions (\ref{kinkq}) and
(\ref{kinkint}) for the induced fermion number are much more 
complicated than the $T=0$ result (\ref{nzero}). Nevertheless, $\langle
N\rangle_{\rm T}$ in (\ref{kinkq}) and (\ref{kinkint}) is still topological in
the sense that it only depends on the kink background through the asymptotic
value $\hat{\theta}={\rm arctan}(\hat{\phi}_2/m)$. Other details of
the kink shape do not matter. We will see in the next subsection that this is
{\it not} true in the sigma model case (\ref{sigma}). It is also interesting
to note that the angular nature of the parameter $\hat{\theta}$ is clearly
manifest in the finite temperature expressions (\ref{kinkq}) and
(\ref{kinkint}), while it is not as obvious from looking at the zero
temperature expression (\ref{nzero}). A similar observation applies
for the finite temperature induced fermion number in $3+1$
dimensions for fermions in the presence of a Dirac monopole
\cite{cp,monopole}, and for fermions in the presence of a static `t
Hooft-Polyakov monopole \cite{goldhaber,ad}.

\subsection{Finite T Fermion Number in the Sigma Model Case}

In the sigma model case (\ref{sigma}), the Callias index theorem result
(\ref{index}) for the even part of the resolvent does not apply. There
is no general expression for the even part of the resolvent. This is
because in the sigma model case
\begin{equation}
H^2=\left(\matrix{-\nabla^2+m^2&0\cr 0&  
-\nabla^2+m^2}\right)+m\, \theta^\prime\, \left(\matrix{\cos\theta& 
-i \sin\theta\cr i \sin\theta& -\cos\theta}\right)
\label{hssigma}
\end{equation}
which is clearly not of the diagonal isospectral form in (\ref{hskink}).
Thus, another approach is needed to evaluate the resolvent. In \cite{ad},
the derivative expansion was used to evaluate the even part of the
resolvent as an expansion in powers of $\theta^\prime$ and its
derivatives. The derivative expansion \cite{ian,ad} assumes that
$\theta^\prime\ll m$ : the spatial derivatives of the background fields are
assumed small compared to the fermion mass scale $m$. In other words,
the background chiral field $\theta(x)$ is assumed to
be slowly varying on the scale of the fermion Compton wavelength. To
next-to-leading order in the derivative expansion for the sigma model
case, one finds \cite{ad}
\begin{eqnarray}
\left[ \tr\left(\frac{1}{H-z}\right)\right]_{\rm even}=
-\frac{m^2}{2(m^2-z^2)^{3/2}}\int dx\, \theta^\prime & -& 
\frac{m^2}{8(m^2-z^2)^{5/2}}
\int dx\, \theta^{\prime\prime\prime}-
\frac{m^2(4z^2+m^2)}{16 (m^2-z^2)^{7/2}} \int dx\, (\theta^\prime)^3 +\dots
\label{third}
\end{eqnarray}
where the dots refer to terms involving five or more derivatives. The
middle term vanishes since $\int dx\, \theta^{\prime\prime\prime}=0$ if
$\theta(x)$ has a kink-like shape. Inserting this approximate expression
into (\ref{even}) we obtain:
\begin{eqnarray}
\langle N\rangle_{\rm T}=\frac{1}{\pi}\left(\frac{m\beta}{\pi}\right)^2
\left\{\sum_{n=0}^\infty 
\frac{1}{[(2n+1)^2+(\frac{m\beta}{\pi})^2]^{3/2}}\, 
\int dx\, 
\theta^\prime + \frac{\beta^2}{8\pi^2}\,
\sum_{n=0}^\infty {[-4(2n+1)^2+(\frac{m\beta}{\pi})^2] \over
[(2n+1)^2+(\frac{m\beta}{\pi})^2]^{7/2}}\, \int dx\,
(\theta^\prime)^3+\dots\right\}
\label{nthird}
\end{eqnarray}
The first term is topological and reduces smoothly to the zero temperature
result (\ref{nzero}) as $T\to 0$.  The second term involves $\int
dx\,(\theta^\prime)^3$, which is clearly {\bf not} topological : it depends on the actual 
{\it shape} of the chiral field $\theta(x)$, not just its asymptotic value 
$\hat{\theta}$. This is still consistent (to this order) with the topological
nature of the zero temperature induced fermion number (\ref{nzero}), because
the energy trace prefactor multiplying the $\int dx\,(\theta^\prime)^3$ term
in (\ref{nthird}) vanishes at $T=0$. 

Higher orders in the derivative expansion (\ref{third}) can be developed
systematically, although it becomes somewhat tedious to enumerate all the
different terms at high orders. However, a remarkable
simplification occurs \cite{ad} in the low temperature limit where $T\ll m$.
The leading low $T$ terms at each order of the derivative expansion have the
simple form :
\begin{equation}
\langle N\rangle^{(2l-1)}_{\rm T}\sim \delta_{l,1}\,\frac{1}{2\pi}\int dx\, \theta^\prime
-\sqrt{\frac{2mT}{\pi}}\, e^{-m/T}\, \frac{1}{(2l-1)!}\,
 \int dx\,  \left(\frac{\theta^\prime}{2T}\right)^{2l-1}+\dots 
\label{asymptotic}
\end{equation}
Thus, in the low temperature limit, we can resum the {\it entire}
derivative  expansion, to obtain the induced
fermion number in the sigma model case (\ref{sigma}) :
\begin{equation}
\langle N\rangle_{\rm T}=\frac{1}{2\pi}\int_{-\infty}^\infty dx\,
\theta^\prime -
\sqrt{\frac{2mT}{\pi}}\,e^{-m/T}\,\int_{-\infty}^\infty dx\,  
{\rm sinh}\left(\frac{\theta^\prime}{2T}\right)+\dots
\label{resum}
\end{equation}
where the dots refer to power-law subleading terms for $T\ll m$.

Note that $\langle N\rangle_{\rm T}$ in (\ref{resum}) naturally splits into a
temperature independent topological term, and a temperature dependent
nontopological term. This has a simple physical explanation \cite{ad}.
First, observe that the chiral sigma model background acts like a spatially
inhomogeneous electric field \cite{rich,wilczek}, as can be seen by making a
local chiral rotation: 
$\psi\to\tilde{\psi}=e^{i\theta \gamma_5/2}\psi$. In
terms of these chirally rotated fields the Lagrangian (\ref{lag}), with 
interaction (\ref{exp}), becomes
\begin{equation}
{\cal L}=i\bar{\tilde{\psi}}\dslash \tilde{\psi}-
m \bar{\tilde{\psi}} \tilde{\psi}
-\bar{\tilde{\psi}}\, \gamma^0\, \frac{\theta^\prime}{2}\,\tilde{\psi}
\label{rotlag}
\end{equation}
Thus, the chiral field $\theta(x)$ acts as an inhomogeneous scalar potential
$A_0(x)=\frac{1}{2}\theta^\prime(x)$, leading to an inhomogeneous electric
field
\begin{equation}
E(x)=-\frac{1}{2}\,\theta^{\prime\prime}(x)
\label{electric}
\end{equation}
This electric field acts on the Dirac sea to polarize the vacuum by
aligning the virtual vacuum dipoles of the Dirac sea, producing a
localized build-up of charge near the kink center. This leads to the
first, topological, term in (\ref{resum}), which is just the familiar zero temperature
result \cite{rich,wilczek}. It is temperature independent as the short-lived virtual
electron-positron dipoles of the Dirac sea do not come to thermal equilibrium. The
second, nontopological, term in (\ref{resum}) arises as the response of the real charges
in the thermal plasma to the spatially inhomogeneous electric field \cite{ad}.
The linear response \cite{lebellac} of the plasma at low temperature to such
an electric field yields an induced fermion number density
\begin{equation}
\rho(x)=\int\frac{dk}{2\pi}\, f(x,k)
\label{linear}
\end{equation}
where $f(x,k)=f_+(x,k)-f_-(x,k)$ is expressed in terms of the particle and antiparticle
Fermi densities. In the derivative expansion limit, the leading effect of
the background is that of an inhomogeneous chemical potential
$\mu(x)=-\frac{1}{2}\,\theta^\prime(x)$. Thus, in this limit, the local Fermi
particle and antiparticle distribution functions are
\begin{equation}
f_\pm(x,k)={1\over e^{\beta (\sqrt{k^2+m^2}\mp \mu(x))}+1}
\label{fermi}
\end{equation}
The resummed derivative expansion expression (\ref{resum}) was obtained in
the low temperature ($T\ll m$) limit. In this limit, we can write
\begin{eqnarray}
f_+(x,k)- f_-(x,k)&\approx & e^{-\beta \sqrt{k^2+m^2}} \left(e^{\beta
\mu(x)}- e^{-\beta\mu(x)}\right)\nonumber\\
&=&-2 e^{-\beta \sqrt{k^2+m^2}}\,
\sinh\left(\frac{\theta^\prime}{2T}\right)
\label{fermiapprox}
\end{eqnarray}
Note that $f_+-f_-$ vanishes smoothly as $T\to 0$ in the derivative expansion regime,
because $\theta^\prime \ll m$. For low temperature, the leading behavior of the k
integral in (\ref{linear}) is
\begin{eqnarray}
\int_{-\infty}^\infty \frac{dk}{2\pi}\, e^{-\beta \sqrt{k^2+m^2}}\sim\,
\sqrt{\frac{mT}{2\pi}}\, e^{-\beta m}+\dots \qquad , \qquad T\to 0
\label{limit}
\end{eqnarray}
Therefore, the plasma linear response contribution (\ref{linear}) to the
induced fermion number gives precisely the second, nontopological, term in
the formula (\ref{resum}) which was obtained by resumming the derivative expansion 
at low temperature.

\section{Thermal Fluctuations of Induced Fermion Number}

\begin{figure}[b]
\includegraphics*{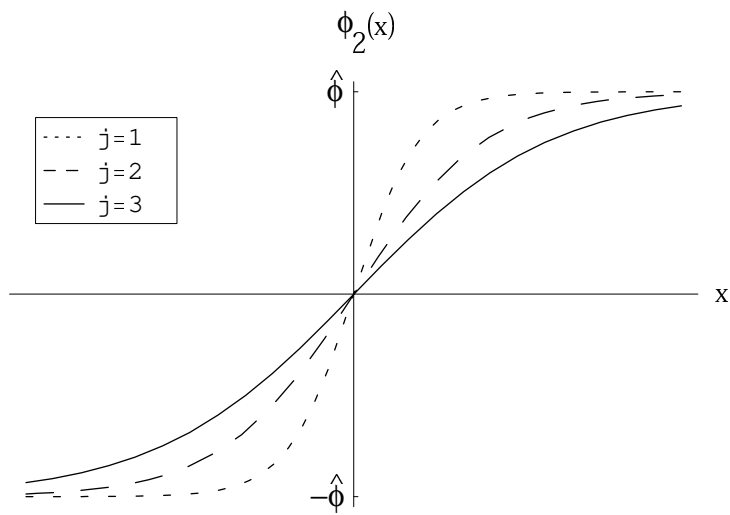}
\caption{Plots of the special kink backgrounds in (\protect{\ref{special}})
for various values of j, illustrating the interpretation of the parameter j
as a scale parameter that characterizes a nontopological aspect of these
kinks.}
\label{f6}
\end{figure}

We now turn to the fluctuation, $(\Delta N)^2_{\rm T}=\langle
N^2\rangle_{\rm T}-\langle N\rangle_{\rm T}^2$, in the induced fermion number
$\langle N\rangle_{\rm T}$. We will show that the fluctuation vanishes at
zero temperature, but is nonzero at finite temperature. Furthermore, the
fluctuation is inherently nontopological, in both the kink and sigma model
cases.

Recall that the fluctuation can be expressed in terms of the partition
function as \cite{pathria}
\begin{eqnarray}
(\Delta N)^2_{\rm T}=\frac{1}{\beta^2}\,
\frac{\partial^2}{\partial \mu^2}\, \left. \log \, {\rm
tr}\,\left(e^{-\beta(H-\mu\, N)}\right)\right]_{\mu=0}
\label{fluctuation}
\end{eqnarray}
Thus, the fluctuation can be expressed in a manner analogous to
the fermion number expressions (\ref{thermalexp}) and (\ref{contour}):
\begin{eqnarray}
(\Delta N)^2_{\rm T} &=& 
\frac{1}{4}\int_{-\infty}^\infty dE\, \sigma(E)\, {\rm sech}^2(\frac{\beta
E}{2}) \nonumber\\
&=&\frac{1}{4}\int_{\cal C}\,\frac{dz}{2\pi i}\,
\tr\left(\frac{1}{H-z}\right) {\rm sech}^2\left(\frac{\beta z}{2}\right)
\label{fluc}
\end{eqnarray}
From a purely computational viewpoint, there is an immediate difference
between (\ref{fluc}) and the fermion number expressions (\ref{thermalexp}) and
(\ref{contour}) : since
${\rm sech}^2(\frac{\beta E}{2})$ is an even function, we now need the
{\it even} part of the spectral function (equivalently, the {\it odd}
part of the resolvent), whereas to compute $\langle N\rangle_{\rm T}$ one
needs the {\it odd} part of the spectral function (equivalently, the {\it
even} part of the resolvent).

\subsection{Fluctuations in the Kink Case}

The Callias index theorem result (\ref{index}) gives the exact form of
the even part of the resolvent, but tells us nothing about the odd part of
the resolvent. Thus, in the kink case, we should no longer expect
an exact topological result for the fluctuation $(\Delta N)^2_{\rm T}$.
Indeed, for a general kink background, an approximation is needed to
compute the odd part of the resolvent. Instead, here we will proceed by
considering a special two-parameter family of kink backgrounds for which the
both the even and odd parts of the resolvent {\it can} be computed exactly.
Consider
\begin{eqnarray}
\phi_2(x)=\hat{\phi}\, \tanh\left(\frac{\hat{\phi}}{j}\, x\right)
\label{special}
\end{eqnarray}
where $j=1,2,3,\dots$ is an integer. The parameter $\hat{\phi}$ represents
the asymptotic value of the kink, while $j$ determines the {\it scale} of
the kink, as illustrated in Fig 6. While not completely general, the
family of kinks in (\ref{special}) is sufficiently general to distinguish
between the topological and nontopological effects. For these special kinks,
the isospectral potentials appearing in the square (\ref{hskink}) of the Dirac
hamiltonian are 
\begin{eqnarray}
\phi_2^2\pm \phi_2^\prime=\hat{\phi}^2\, \left[ 1-\frac{(j\mp 1)}{j} \,
{\rm sech}^2 \left(\frac{\hat{\phi} }{j}\, x\right)\right]
\label{jiso}
\end{eqnarray}
which are exactly solvable
reflectionless P\"oschl-Teller potentials, for which the resolvents are known
in closed form, as is most easily derived from the exact phase shifts 
\cite{morse,landau,noah}. In fact, for the family of kink fields $\phi_2(x)$ in
(\ref{special}), the associated Schr\"odinger resolvents are :
\begin{eqnarray} 
\tr\left({1\over -\nabla^2+\phi_2^2+
\phi_2^\prime+m^2-z^2}\right)&=&{\hat{\phi}\over
\sqrt{m^2+\hat{\phi}^2-z^2}}\, \sum_{l=1}^{j-1} {l/j \over
\left(m^2+\hat{\phi}^2(1-\frac{l^2}{j^2})-z^2\right)} 
\label{isoplus}\\
\tr\left({1\over -\nabla^2+\phi_2^2-
\phi_2^\prime+m^2-z^2}\right)&=&{\hat{\phi}\over
\sqrt{m^2+\hat{\phi}^2-z^2}}\, \sum_{l=1}^{j} {l/j \over
\left(m^2+\hat{\phi}^2(1-\frac{l^2}{j^2})-z^2\right)} 
\label{isominus}
\end{eqnarray}
Note that the only difference between (\ref{isoplus}) and
(\ref{isominus}) is the $l=j$ term in the sum, which is present in
(\ref{isominus}) but not in (\ref{isoplus}). Since the even part of
the Dirac resolvent (\ref{susy}) involves the {\it difference} of these two 
isospectral Schr\"odinger resolvents, we see that this difference is indeed
independent of the scale $j$, and confirms the topological Callias index
theorem result (\ref{index}) for this case. On the other hand, the odd part
of the Dirac resolvent depends on both the asymptotic value $\hat{\phi}$ and
the scale
$j$:
\begin{eqnarray}
\left[\tr\left(\frac{1}{H-z}\right)\right]_{\rm odd} &=&
{|\hat{\phi}|\, z\over
(m^2-z^2)\sqrt{m^2+\hat{\phi}^2-z^2}}+{2|\hat{\phi}|\, z\over
\sqrt{m^2+\hat{\phi}^2-z^2}}\, \sum_{l=1}^{j-1} {l/j \over
\left(m^2+\hat{\phi}^2(1-\frac{l^2}{j^2})-z^2\right)} 
\label{oddodd}
\end{eqnarray}
The modulus $|\hat{\phi}|$ appears in the Dirac resolvent (\ref{oddodd})
because the complex Dirac energy $z$ is the square root of the Schr\"odinger
energy parameter $z^2$ appearing in (\ref{isoplus}) and (\ref{isominus}), and
so we must be careful about the sheet structure in the complex $z$ plane.
For example, from the Dirac Hamiltonian (\ref{ham}) it follows that if
$\hat{\phi}>0$ then there is an unpaired bound state
$\psi=\left(\matrix{0 \cr \exp(-\int \phi_2)}\right)$ at
$E=-m$, while if 
$\hat{\phi}<0$ then there is an unpaired bound state
$\psi=\left(\matrix{\exp(\int \phi_2)\cr 0}\right)$ at $E=+m$. This fact
is encapsulated in the Dirac resolvent (\ref{index}) and (\ref{oddodd}),
and similar arguments apply for the other bound states and for the
continuum cuts. 

\begin{figure}
\includegraphics*{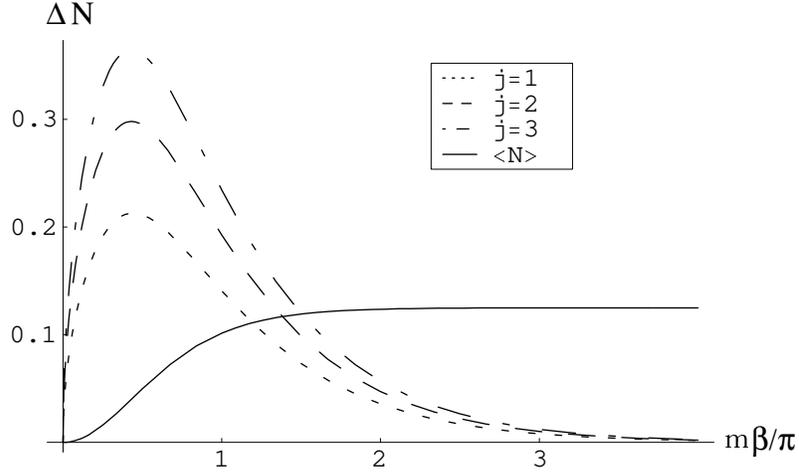}
\caption{Plots of the fluctuation $\Delta N$, given in (\protect{\ref{kfs}})
and (\protect{\ref{kfi}}), for the special kink backgrounds in
(\protect{\ref{special}}). The plots show the dependence of $\Delta N$ on the
inverse temperature $\frac{m\beta}{\pi}$, for various values of the kink
scale j, as indicated. This plot is for $\hat{\theta}=\pi/8$. The solid line 
shows the corresponding value of the fermion number $\langle N\rangle$, which
is topological and therefore independent of the scale j. Notice that
$\Delta N$ vanishes exponentially as the temperature approaches zero, while 
$\langle N\rangle$ saturates to its zero temperature value of $1/8$. These
plots show the dependence of the fluctuation $\Delta N$ on the scale j,
indicating that $\Delta N$ is nontopological.}
\label{f7}
\end{figure}

Given this result (\ref{oddodd}) for the odd part of the resolvent, the
contour integral expression (\ref{fluc}) for the fluctuation yields a
summation expression:
\begin{eqnarray}
(\Delta N)_{\rm T}^2&=& \frac{2m\beta |\sin\hat{\theta}|}{\pi^3}
\, \sum_{n=0}^\infty\left\{ {2(2n+1)^4
\cos^2\hat{\theta}-(\frac{m\beta}{\pi})^2\left[(\frac{m\beta}{\pi})^2-
(2n+1)^2\right]\over \left[(\frac{m\beta}{\pi})^2+(2n+1)^2\right]^2 
\left[(\frac{m\beta}{\pi})^2+(2n+1)^2\cos^2\hat{\theta}\right]^{3/2}}\right.
\nonumber\\
&& \left. +2 \sum_{l=1}^{j-1}\left(\frac{l}{j}\right)\,  {2(2n+1)^4
\cos^2\hat{\theta}-(\frac{m\beta}{\pi})^2\left[(\frac{m\beta}{\pi})^2-
(2n+1)^2\right]
-(\frac{m\beta}{\pi})^4(1-\frac{l^2}{j^2})\tan^2\hat{\theta} \over
\left[(\frac{m\beta}{\pi})^2\left(1+(1-\frac{l^2}{j^2})\tan^2\hat{\theta}\right)
+(2n+1)^2\right]^2 \left[(\frac{m\beta}{\pi})^2+
(2n+1)^2\cos^2\hat{\theta} \right]^{3/2} } \right\}
\label{kfs}
\end{eqnarray}
Alternatively, there is an equivalent integral representation expression:
\begin{eqnarray}
(\Delta N)_{\rm T}^2&=&\frac{1}{4}\left\{ {\rm sech}^2(\frac{m\beta}{2})
-\frac{2|\sin\hat{\theta}|}{\pi}\int_1^\infty du\, {u\, {\rm
sech}^2\left(\frac{m\beta}{2}{\rm sec}\hat{\theta}\, u\right) \over
(u^2-\cos^2\hat{\theta})\sqrt{u^2-1}}\right. \nonumber\\ 
&& \hskip -4cm +2  \sum_{l=1}^{j-1} \left.
\left[ 
{\rm sech}^2\left(\frac{m\beta}{2}
\sqrt{1+(1-\frac{l^2}{j^2})\tan^2\hat{\theta}}\right)
- \frac{2|\sin\hat{\theta}|}{\pi} \left(\frac{l}{j}\right)\, \int_1^\infty du\, 
{u\, {\rm sech}^2\left(\frac{m\beta}{2}{\rm sec}\hat{\theta}\, u\right) \over
\left(u^2-(1-\frac{l^2}{j^2}\sin^2\hat{\theta}) \right) \sqrt{u^2-1}} \right] \right\}
\label{kfi}
\end{eqnarray}
The fluctuations are plotted in Fig. 7 for 
$\hat{\theta}=\frac{\pi}{8}$, and various values of the scale parameter $j$.
It is clear that the fluctuation vanishes exponentially at zero
temperature ($m\beta\to\infty$). It is also clear that the fluctuation is
nontopological, as it depends on the scale parameter $j$.  At low
temperature, the leading behavior is independent of the scale $j$
\begin{eqnarray}
(\Delta N)_{\rm T}^2\sim e^{-m\beta} +\dots,
\label{leadingf}
\end{eqnarray}
but the subleading exponential corrections are $j$ dependent, as can be seen
from the integral expression (\ref{kfi}) and from the plots in Fig. 7. 

\subsection{Fluctuations in Sigma Model Case}

In the sigma model case, there is no exact expression for either the odd
or the even part of the Dirac resolvent. However, applying the derivative
expansion as in \cite{ad}, we find that in the low temperature ($T\ll m$)
limit the dominant contribution comes from terms involving (even) powers of
$\theta^\prime(x)$. These can be evaluated and resummed to all orders of
the derivative expansion, to yield the leading low temperature result :
\begin{eqnarray}
(\Delta N)_{\rm T}^2\,\sim\, 2\sqrt{\frac{2mT}{\pi}}\,e^{-m/T}\, \int
dx\,
 \sinh^2\left(\frac{\theta^\prime}{4 T}\right)+\dots
\label{fsm}
\end{eqnarray}
This expression has a smooth low temperature limit in the
derivative expansion regime where $\theta^\prime\ll m$. Indeed,  the
fluctuation  (\ref{fsm}) vanishes as $T\to 0$. Furthermore, at any finite
temperature, the fluctuation (\ref{fsm}) is clearly nontopological. In the
next Section we will see that this expression for the fluctuation has a
simple plasma interpretation, analogous to the linear response explanation
(\ref{linear}) -- (\ref{limit}) of the induced fermion number (\ref{resum}) in the sigma
model case.

\section{Physical Interpretation}

\begin{figure}[b]
\includegraphics[scale=0.75]{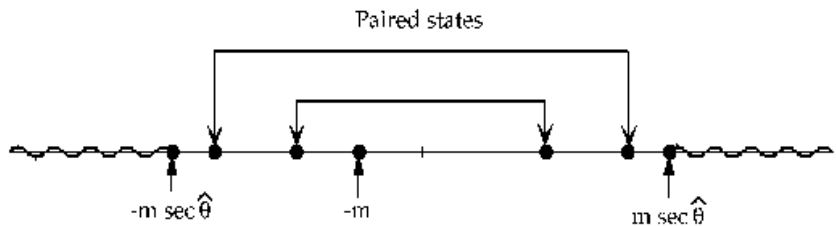}
\caption{The general structure of the Dirac spectrum for a kink
background. There is a single unpaired bound state at $E=-m$. (This
assumes $\hat{\theta}>0$. If $\hat{\theta}<0$, then the unpaired bound
state would be at $E=+m$.) There are continuum thresholds at $E=\pm
m\,{\rm sec}\hat{\theta}=\pm \sqrt{m^2+\hat{\phi}_2^2}$. There may or may
not be additional bound states with energies in the range $m< |E|<m\, 
{\rm sec}\hat{\theta}$, but if these states are present they are
necessarily paired in $\pm E$ pairs.}
\label{f8}
\end{figure}

\begin{figure}[t]
\includegraphics[scale=0.75]{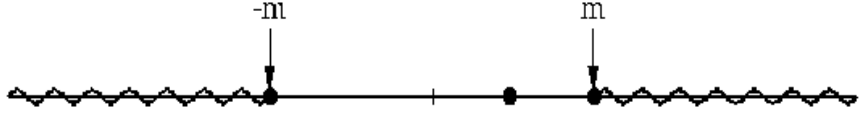}
\caption{The structure of the Dirac spectrum for a sigma model
background. The only general thing that can be said is that the continuum
thresholds appear at $E=\pm m$. There is no particular symmetry to the
spectrum, and there may or may not be one or more bound states in the gap.
This depends sensitively on the detailed form of the angular field 
$\theta(x)$.}
\label{f9}
\end{figure}

In this Section we give a simple physical interpretation of the results
of the previous Sections for the induced fermion number and its fluctuations,
in terms of the spectrum of the fermions in the background field. Let us
first summarize the results. At zero temperature, the induced fermion number
$\langle N\rangle_0$ is always topological, and has vanishing fluctuation:
$(\Delta N)_0=0$. At finite temperature, $\langle N\rangle_{\rm T}$ is
topological for a kink background, but nontopological for a sigma model
background. For both the kink and sigma model cases, the fluctuation is
non-vanishing, $(\Delta N)_{\rm T}\neq 0$, at nonzero temperature, and 
$(\Delta N)_{\rm T}$ is nontopological. The vanishing of the fluctuations at
zero temperature is in agreement with previous arguments that the zero
temperature induced fermion number is a sharp observable
\cite{kivelson,bell,jackiw,frishman,charge}. At finite temperature, the
nonvanishing fluctuation indicates that the induced fermion number is no
longer a sharp observable. Rather, it is a thermal expectation value that
mixes contributions from different states in the fermion Hilbert space. This
is somewhat analogous to the recent observation of Jackiw {\it et al}
\cite{helium} that fractional electron number in liquid helium bubbles is not
sharp, due to state mixing.

In the sigma model case, the resummed derivative expansion result
(\ref{resum}) for $\langle N\rangle_{\rm T}$ suggested a natural separation of
the finite temperature induced fermion number $\langle N\rangle_{\rm T}$ into
a zero temperature topological piece $\langle N\rangle_0$ and a
nontopological piece that represents the plasma response to the background
field (in this case, an inhomogeneous electric field). This separation is, in
fact, a general feature of the induced fermion number:
$\langle N\rangle_{\rm T}$ splits into a topological  piece
$\langle N\rangle_0$ that arises due to vacuum polarization effects (and
only involves the ground state of the second quantized Hilbert space), and a
nontopological piece that can be interpreted in terms of the thermal 
occupation numbers of nonvacuum states. The topological piece is temperature
independent, while the nontopological piece is temperature dependent.
To see how this separation arises, note the simple identity
\begin{eqnarray}
\tanh\left(\frac{\beta E}{2}\right)=1-2 n(E) 
\label{tanh}
\end{eqnarray}
where the Fermi occupation number distribution is
\begin{eqnarray}
n( E)=\frac{1}{e^{\beta E}+1}
\label{number}
\end{eqnarray}
Then the finite temperature induced fermion number (\ref{thermalexp})
separates as
\begin{eqnarray}
\langle N\rangle_{\rm T}&=&-\frac{1}{2}\int_{-\infty}^\infty dE\, \sigma(E)\,
{\rm sign}(E)+\int_{-\infty}^\infty dE\, \sigma(E)\, {\rm
sign}(E)\, n(|E|)
\label{split}
\end{eqnarray}
The first term in (\ref{split}) is just the zero temperature fermion number
(\ref{asymmetry}), and is topological.  However, the fact that the spectral
asymmetry integral in (\ref{asymmetry}) is always topological does not
necessarily mean that $\sigma_{\rm odd}(E)$ is itself topological. The
second term in (\ref{split}) is generically nontopological due to the
weighting of the integral with the Fermi distribution factor. However, in
the special case of a kink background,
$\sigma_{\rm odd}(E)$ is itself topological [see the Callias index theorem
result (\ref{index})]. Thus the temperature-dependent
second term in (\ref{split}) is also topological in the kink case, even
though the integral involves the Fermi factor.

This separation can be understood in terms of the Dirac spectrum of the
fermions in the given background. For example, in the kink case, the
Sommerfeld-Watson integral representation (\ref{kinkint}) for the
induced fermion number can be re-expressed using (\ref{int}), (\ref{tanh}) and
(\ref{split}) as:
\begin{eqnarray}
\langle N\rangle_{\rm T}=\langle N\rangle_0 -{\rm sign}(\hat{\theta})\, n(m)
+\frac{2 m^2\tan\hat{\theta}}{\pi}
\int_{m{\rm sec}\hat{\theta}}^\infty dE\, \frac{n(
E)}{(E^2-m^2)\sqrt{E^2-m^2{\rm sec}^2\hat{\theta}}}
\label{kinksp}
\end{eqnarray}
For {\it any} kink background (with $\hat{\theta}$ positive), the
spectrum of the Dirac Hamiltonian (\ref{ham}) always has the form shown in
Fig. 8. There are continuum states for
$|E|>\sqrt{m^2+\hat{\phi}_2^2}=m\, {\rm sec}\,\hat{\theta}$, where we
recall that 
$\hat{\theta}\equiv {\rm arctan}(\hat{\phi_2}/m)$, and $\hat{\phi_2}$ is
the asymptotic value of the kink field at $x=+\infty$.  In addition,
there is always a bound state at $E=-{\rm sign}(\hat{\theta})\,m$:
\begin{eqnarray}
\psi_b&=&\left(\matrix{0\cr \exp[-\int^x \phi_2]}\right)\quad {\rm
with\,\,energy} \,\, E=-m,\,\, {\rm if}\,\,
\hat{\theta}>0\nonumber\\
\psi_b&=&\left(\matrix{\exp[\int^x \phi_2]\cr
0}\right)\quad {\rm with\,\,energy} \,\, E=+m,\,\, {\rm if}\,\,
\hat{\theta}<0
\label{zero}
\end{eqnarray}
There may or may not be additional bound states with
$m<|E|<m\,{\rm sec}\,\hat{\theta}$. If these additional bound states are
present they necessarily occur in $\pm E$ pairs, because of the quantum
mechanical SUSY of the Dirac Hamiltonian (\ref{ham}) in the kink case.
Thus, the contributions of these paired bound states to the second
integral in (\ref{split})   cancel in pairs, while the unpaired bound
state at
$E=-{\rm sign}(\hat{\theta})\,m$ leads to the
$-{\rm sign}(\hat{\theta})\,n( m)$ term in (\ref{kinksp}). The remaining
integral over the continuum states leads to the integral term in
(\ref{kinksp}), as an integral of the Fermi factor $n(|E|)$ over the
continuum beginning at the threshold energy $m{\rm sec}\hat{\theta}$, with
the integrand involving the odd part of the spectral function derived from
the Callias index theorem result (\ref{index}).  

In the sigma model case, the Dirac spectrum is very different. As
discussed in Section III.B, the chiral field $\theta(x)$ is equivalent to
a spatially inhomogeneous electric field
$E(x)=-\frac{1}{2}\theta^{\prime\prime}(x)$ for fermions of mass $m$. For a
slowly varying background, with $\theta^\prime\ll m$, the leading order
effect on the spectrum is that of a local chemical potential
$\mu(x)=-\frac{1}{2}
\theta^\prime(x)$. Thus, we can approximate (\ref{split}) as
\begin{eqnarray}
\langle N\rangle_{\rm T}&\approx &\langle N\rangle_0+\int_{-\infty}^\infty dx
\int_{-\infty}^\infty
\frac{dk}{2\pi}\,\left(\frac{1}{e^{\beta(\sqrt{k^2+m^2}+\theta^\prime/2)}+1}-
 \frac{1}{e^{\beta(\sqrt{k^2+m^2}-\theta^\prime/2)}+1}\right)\nonumber\\
&\sim &\langle N\rangle_0-\sqrt{\frac{2mT}{\pi}}\,
e^{-m/T}\, \int_{-\infty}^\infty dx\,
\sinh\left(\frac{\theta^\prime}{2T}\right)\quad ,\quad {\rm as}\quad T\to 0
\label{plasma}
\end{eqnarray}
which agrees precisely with the low temperature resummed derivative expansion
result (\ref{resum}), and which agrees with the plasma linear response 
result as explained in Section III.B.

From numerical analysis of the Dirac equation for a sigma model
background, the Dirac spectrum has the form shown in Fig. 9. Note that,
unlike the kink case spectrum depicted in Fig. 8, there is no particular
symmetry of the spectrum. The only general thing we can say is that the
continuum thresholds are at $E=\pm m$. There may or may not be one or more
bound states for $|E|<m$. The existence of (and the precise location of) bound
states is highly sensitive to the actual {\it shape} of the $\theta(x)$. As
there is no special symmetry in the spectrum, there is nothing general that
can be said about the odd part of the spectral function. Nevertheless, the
first term in (\ref{split}), which is $\langle N\rangle_0$, yields a topological result
since the integral can be written, using Levinson's theorem, in terms of the phase shift
at infinity, which is topological \cite{yamagishi,boy,mike,jaffe2}. This
application of Levinson's theorem doesn't work for the second term in
(\ref{split}) because of the Fermi factor. This is another way to understand
why the finite temperature induced fermion number is generically
nontopological. At low temperature, we see from (\ref{split}) that the
dominant correction to the zero temperature fermion number is determined by
the bound state energy with the lowest magnitude:
\begin{eqnarray}
\langle N\rangle_{\rm T}\sim \langle N\rangle_0+{\rm sign}(E_{\rm min})\,
e^{-\beta\, |E_{\rm min}|}+\dots \quad , \quad {\rm as}\quad T\to 0
\label{lowtsigma}
\end{eqnarray}
Since the location of $E_{\rm min}$ is sensitive to the detailed shape of
$\theta(x)$, this shows that $\langle N\rangle_{\rm T}$ is generically
nontopological. If there is no bound state, then the leading correction
comes from the threshold and is $\sim e^{-m\beta}$, with nontopolgical
subleading corrections. 

The fluctuation (\ref{fluc}) can also be expressed in terms of the Fermi
distribution functions $n(E)$ in (\ref{number}). Noting the simple identity
\begin{eqnarray}
\frac{1}{4}\,{\rm sech}^2\left(\frac{\beta
E}{2}\right)=n(E)\left(1-n(E)\right)
\label{sech}
\end{eqnarray}
we see that (\ref{fluc}) can be written as
\begin{eqnarray}
(\Delta N)^2_{\rm T}=\int_{-\infty}^\infty dE\, \sigma(E) \,
n(E)\,\left(1-n(E)\right)
\label{fsplit}
\end{eqnarray}
This formula is natural, since it is well known \cite{pathria} that for
noninteracting fermions, the fluctuation in the occupation number of the
state with energy $E$ is $\langle n^2_E\rangle-\langle n_E\rangle^2=
\langle n_E\rangle \left(1-\langle n_E\rangle \right)$. In this paper we are
considering the fermions to be in a fixed static background, so we are simply
populating the single-particle energy levels of the corresponding Dirac
Hamiltonian with noninteracting fermions, according to Fermi-Dirac
statistics. This explains why the fluctuation vanishes at zero temperature,
as the $n(1-n)$ factors vanish exponentially fast. It also explains why the
fluctuation is nontopological, since the integral in (\ref{fsplit}) requires
detailed knowledge of the spectrum. 

For the family of special kink backgrounds in (\ref{special}), the
integral representation result (\ref{kfi}) for the fluctuation can be
re-expressed in terms of the Fermi occupation numbers of the states in the
spectrum:
\begin{eqnarray}
(\Delta N)^2_{\rm T}&=&
n(m)\left(1-n(m)\right)+2\sum_{l=1}^{j-1}n(E_l)\left(1-n(E_l)\right)
-\frac{2 m|\tan\hat{\theta}|}{\pi}
\int_{m{\rm sec}\hat{\theta}}^\infty dE\,
\frac{E\,n(E)\left(1-n(E)\right)}{(E^2-m^2)\sqrt{E^2-m^2{\rm
sec}^2\hat{\theta}}}
\nonumber\\
&&-\frac{4 m|\tan\hat{\theta}|}{\pi}\sum_{l=1}^{j-1}\,
\left(\frac{l}{j}\right)
\int_{m{\rm sec}\hat{\theta}}^\infty dE\,
\frac{E\,n(E)\left(1-n(E)\right)}{(E^2-E_l^2)\sqrt{E^2-m^2{\rm
sec}^2\hat{\theta}}}
\label{kfsplit}
\end{eqnarray}
where the paired bound state energies $\pm E_l$ for the special kink
backgrounds (\ref{special}) are at
\begin{eqnarray}
E_l=m\sqrt{1+(1-\frac{l^2}{j^2})\tan^2\hat{\theta}}\qquad ,\qquad
l=1,2,\dots , (j-1)
\label{bound}
\end{eqnarray}
This expression (\ref{kfsplit}) can be interpreted directly in terms of the
general expression (\ref{fsplit}) : we can easily identify the separate
contributions from the unpaired bound state at $E=-m$, from the paired bound
states at $E=\pm E_l$, and the contributions from the continuum cuts. Since
the integral in (\ref{fsplit}) involves the even part of the spectral
function, the symmetry of the spectrum indicated in Fig. 8 does not lead to
any particular simplification for the fluctuation, as there are no
cancellations. Hence, $(\Delta N)_{\rm T}^2$ is nontopological, as it is manifestly
dependent on the scale parameter $j$.

In the sigma model case, we can interpret the equation (\ref{fsplit}) as
follows. As discussed previously, in the derivative expansion,
$\theta^\prime\ll m$, limit, the leading effect of the background is that of
a local chemical potential $\mu(x)=-\frac{1}{2}\theta^\prime(x)$, in which
case the local Fermi distribution factors are given by (\ref{fermi}). Thus,
the fluctuation can be computed from (\ref{fsplit}) as
\begin{eqnarray}
(\Delta N)^2_{\rm T}&\approx &\int dx \int_{-\infty}^\infty \frac{dk}{2\pi}
\left[f_+(1-f_+)+f_-(1-f_-)-2 f_0(1-f_0)\right]\nonumber\\
&= &\int dx \int_{-\infty}^\infty \frac{dk}{2\pi}
\left[ n(\sqrt{k^2+m^2}-\mu(x))\left(1-n(\sqrt{k^2+m^2}-\mu(x))\right)\right.
\nonumber\\
&&\left.+n(\sqrt{k^2+m^2}+\mu(x))\left(1-n(\sqrt{k^2+m^2}+\mu(x))\right)-
2 n(\sqrt{k^2+m^2})\left(1-n(\sqrt{k^2+m^2})\right)\right]
\label{fsigma}
\end{eqnarray}
where the last term is the subtraction of the free case with 
$f_0=n(\sqrt{k^2+m^2})$. At low temperature, (\ref{fsigma}) gives the leading 
contribution
\begin{eqnarray}
(\Delta N)^2_{\rm T}&\approx &\int dx \int_{-\infty}^\infty \frac{dk}{2\pi}\,
e^{-\beta\sqrt{k^2+m^2}} \, 
\left[e^{-\beta\theta^\prime/2}+e^{\beta\theta^\prime/2}-2\right] \nonumber\\
&\sim & 4\sqrt{\frac{mT}{2\pi}}\, e^{-m\beta}\, \int dx \,
\sinh^2\left(\frac{\theta^\prime}{4T}\right)
\label{fs}
\end{eqnarray}
where we have used the low temperature behavior of the $k$ integral in
(\ref{limit}). This expression (\ref{fs}) agrees precisely with the resummed
derivative expansion result (\ref{fsm}) for the fluctuation.

\section{Conclusions}

To conclude, we reiterate that the finite temperature induced fermion number
$\langle N\rangle_{\rm T}$ is generically nontopological, and is not a
sharp observable. The induced fermion number naturally splits into a
topological temperature-independent piece that represents the effect of
vacuum polarization on the Dirac sea, and a nontopological
temperature-dependent piece that represents the thermal
population of the available states in the fermion spectrum, weighted with the
appropriate Fermi-Dirac factors. As the temperature approaches zero, the
nontopological terms vanish exponentially fast, and we regain smoothly the
familiar topological results at zero temperature. But at nonzero temperature,
the nontopological contribution is more sensitive to the details of the
spectrum, and so is generically nontopological.  This is illustrated
explicitly by a derivative expansion calculation, which is resummed to all
orders at low temperature, for the case of a sigma model background. The
induced fermion number $\langle N\rangle_{\rm T}$ only depends on the {\it
odd} part of the spectral function. This means the case of a kink background
is special because the Dirac spectrum for a kink background is remarkably
symmetric, and the odd part of the spectral function is itself a topological
quantity, in the sense that it only depends on the asymptotic value of the
kink field, rather than on the full details of the shape of the kink field.
Thus, the kink case is non-generic, and the finite temperature induced
fermion number is actually topological, albeit a complicated function of the
temperature and the asymptotic value of the kink field.  On the other hand,
the fluctuation, $(\Delta N)_{\rm T}$, is determined by the
even part of the spectral function, for which no general topological
information is known. Thus, the fluctuations are always nontopological at
finite temperature. We illustrated this explicitly with an exact solution for
a special class of kink backgrounds, and with a derivative expansion
calculation for the sigma model case. The fluctuation vanishes at zero
temperature, which shows that the induced fermion number is a sharp
eigenvalue at zero temperature, in agreement with the arguments of
\cite{kivelson,bell,jackiw,frishman,charge}. But at finite temperature, the
nonvanishing of the fluctuations indicates that the induced fermion number is
not a sharp eigenvalue at nonzero temperature.  The source of the fluctuation
is clearly  the dispersion introduced by mixing states in the thermal
average. An analogous example of an induced fermion number expectation value
that is not sharp has been discussed recently in \cite{helium}.

Some of these results generalize immediately to higher dimensional cases.
For example, the dependence of $\langle N\rangle_{\rm T}$ on just the odd
part of the spectrum means that if the background has the same SUSY
isospectral features as the kink case, then the finite temperature
induced fermion number will be topological. This was in fact found to be
true in \cite{ad} where the
$3+1$ dimensional induced fermion number in the presence of a static $SU(2)$
't Hooft-Polyakov monopole was shown to have exactly the same form as in the
$1+1$ dimensional kink case, with the identification of the asymptotic value
of the kink with the asymptotic value of the magnitude of the monopole's Higgs
field. This gives an interesting finite temperature remnant of the quantum mechanical
SUSY present in the Dirac spectrum. Conversely,
without such a special symmetry of the Dirac spectrum, the finite temperature
induced fermion number will be generically nontopological. So, for example,
for a chiral $SU(2)$ sigma model with a Skyrme background in $3+1$
dimensions, it is well known from numerical work that the Dirac energy
spectrum of the fermions is not symmetric, and the energy of a possible
bound state (and indeed the number of bound states) is highly sensitive to 
the details of the shape of the radial hedgehog field \cite{ripka}. Since 
the leading low temperature contribution to the finite temperature fermion 
number in (\ref{split}) is determined by the lowest magnitude bound state energy
[see (\ref{lowtsigma})], this shows that in this case the finite temperature induced
fermion number is nontopological. Likewise, the fact that the finite temperature
fluctuation $(\Delta N)_{\rm T}$ depends on the even part of the spectral function
indicates that this will be generally nontopological. Finally, the interpretation in
terms of the fluctuations for individual energy levels in the spectrum shows that
the fluctuation is nonzero at finite temperature, but vanishing at zero
temperature. Thus, the induced fermion number is a sharp observable at
zero temperature, but not at finite temperature. In particular, for a
Skyrme background in $3+1$ dimensions, the finite temperature fermion
number is not simply the (topological) winding number of the Skyrme field,
but a much more complicated nontopological object.

\begin{acknowledgments}
We thank the U.S. Department of Energy for support through grant
DE-FG02-92ER40716.00. 
\end{acknowledgments}


\end{document}